\documentclass[twocolumn,prb,aps,showpacs]{revtex4-1}

\usepackage{graphicx}
\usepackage{color}
\usepackage{dcolumn}
\usepackage{amsmath}
\usepackage{units}
\usepackage{amsmath,amssymb}
\usepackage{bibentry}

\newlength{\figwidth} {\tiny }
 \newlength{\figwidthb} %
\newcommand{\CRO}{{Ca$_2$RuO$_4$} } 
\newcommand{\CROns}{Ca$_2$RuO$_4$}

\newcommand{\vc}[1]{\boldsymbol{#1}}

\begin{document}
\title{Higgs mode and its decay in a two dimensional antiferromagnet} 

\author{A. Jain$^{1,2,\dagger}$, M. Krautloher$^{1,\dagger}$, J. Porras$^{1,\dagger}$,  G. H. Ryu$^1$, D. P. Chen$^1$, D. L. Abernathy$^3$, J. T. Park$^4$, A. Ivanov$^5$, J. Chaloupka$^6$, G. Khaliullin$^1$, B. Keimer$^{1,*}$, and B. J. Kim$^{1,7,*}$}


\address{$^1$Max Planck Institute for Solid State Research, Heisenbergstra\ss e 1, D-70569 Stuttgart, Germany} 
\address{$^2$Solid State Physics Division, Bhabha Atomic Research Centre, Mumbai 400085, India}
\address{$^3$Quantum Condensed Matter Division, Oak Ridge National Laboratory, Oak Ridge, Tennessee 37831, USA}
\address{$^4$Heinz Maier-Leibnitz Zentrum, TU M\"unchen, Lichtenbergstra\ss e 1, D-85747 Garching, Germany}
\address{$^5$Institut Laue-Langevin, 6, rue Jules Horowitz, BP 156, 38042 Grenoble Cedex 9, France}
\address{$^6$ Central European Institute of Technology,
Masaryk University, Kotl\'a\v{r}sk\'a 2, 61137 Brno, Czech Republic}
\address{$^7$Department of Physics, Pohang University of Science and Technology, Pohang 790-784, Republic of Korea}

\maketitle



\noindent
{\bf
Condensed-matter analogs of the Higgs boson in particle physics allow insights into its behavior in different symmetries and dimensionalities\cite{pekker_2014}. Evidence for the Higgs mode has been reported in a number of different settings, including ultracold atomic gases\cite{Endres_2012}, disordered superconductors\cite{Sherman_2015}, and dimerized quantum magnets\cite{Ruegg_2008}. However, decay processes of the Higgs mode (which are eminently important in particle physics) have not yet been studied in condensed matter due to the lack of a suitable material system coupled to a direct experimental probe. A quantitative understanding of these processes is particularly important for low-dimensional systems where the Higgs mode decays rapidly and has remained elusive to most experimental probes. Here, we discover and study the Higgs mode in a two-dimensional antiferromagnet using spin-polarized inelastic neutron scattering.  Our spin-wave spectra of \CRO directly reveal a well-defined, dispersive Higgs mode, which quickly decays into transverse Goldstone modes at the antiferromagnetic ordering wavevector. Through a complete mapping of the transverse modes in the reciprocal space, we uniquely specify the minimal model Hamiltonian and describe the decay process. We thus establish a novel condensed matter platform for research on the dynamics of the Higgs mode.
}

\begin{figure}
\centerline{\includegraphics[width=1\columnwidth,angle=0]{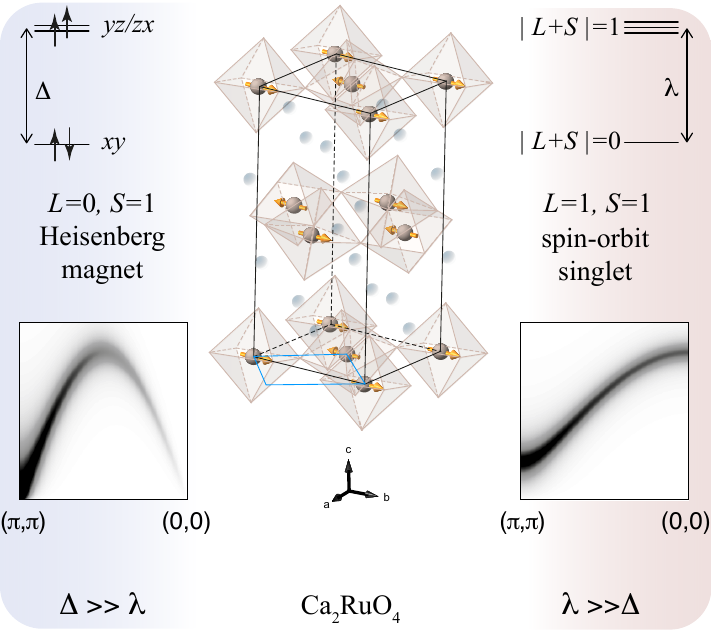}}
\caption{{\bf Crystal, magnetic, and electronic structures of \CROns.} \CRO crystallizes in the orthorhombic {\it Pbca} space group, a distorted variant of the layered perovskite structure with a quasi-two-dimensional square lattice. For clarity, Ca ions are shown as small, light grey balls and oxygen ions are not shown. The distortion involves 2$\%$ compression of the RuO$_6$ octahedra along the $c$-axis, and their rotation about the $c$-axis and titling about an axis that lies in the $ab$ plane\cite{Braden_1998,Friedt2001}. ($\pi$,$\pi$) magnetic order develops below $T_\mathrm{N}$\,$\approx$\,110 K with the moment (orange arrow) aligned approximately along the $b$-axis. The compressive distortion of the RuO$_6$ leads to the splitting $\Delta$ between the orbitals of $xy$ and $yz/zx$ symmetry. If $\Delta$ is much larger than the spin-orbit splitting ($\lambda$), the orbital degrees of freedom are completely quenched and a $S$\,=\,1 Heisenberg magnet is obtained. In the other limit $\lambda$\,$\gg$\,$\Delta$, a non-magnetic singlet ground state is stabilized. These two distinct phases exhibit qualitatively different magnetic excitation spectra. See Figs.~S1 and Fig.~S2 for the evolution of the electronic structure and the spin-wave dispersions between these two limiting cases.}\label{fig:fig1}
\end{figure}

For a system of interacting spins, amplitude fluctuations of the local magnetization---the Higgs mode---can exist as well-defined collective excitations near a quantum critical point (QCP). We consider here a magnetic instability driven by the intra-ionic spin-orbit coupling, which tends toward a nonmagnetic state through complete cancellation of orbital ($L$) and spin ($S$) moments when they are antiparallel and of equal magnitude\cite{Khaliullin_2013,
Meetei_2015}. Specifically, we investigate the magnetic insulator \CROns, a quasi-two-dimensional antiferromagnet \cite{Nakatsuji_1997} with nominally $L$=1 and $S$=1 (Fig.~1). Because the local symmetry around the Ru(IV) ion is very low\cite{Braden_1998,Friedt2001} (having only inversion symmetry), it is widely believed that the orbital moment is completely quenched by the crystalline electric field\cite{Anisimov_2002,Fang2004,Liebsch_2007,Gorelov_2010}, which is dominated by the compressive distortion of the RuO$_6$ octahedra along the $c$-axis (Fig.~1). In the absence of an orbital moment, the nearest-neighbor magnetic exchange interaction is necessarily isotropic. Deviations from this behavior are a sensitive indicator of an unquenched orbital moment. If this moment is sufficiently strong, it can drive \CRO close to a QCP with novel Higgs physics.

\begin{figure}
\centerline{\includegraphics[width=1\columnwidth,angle=0]{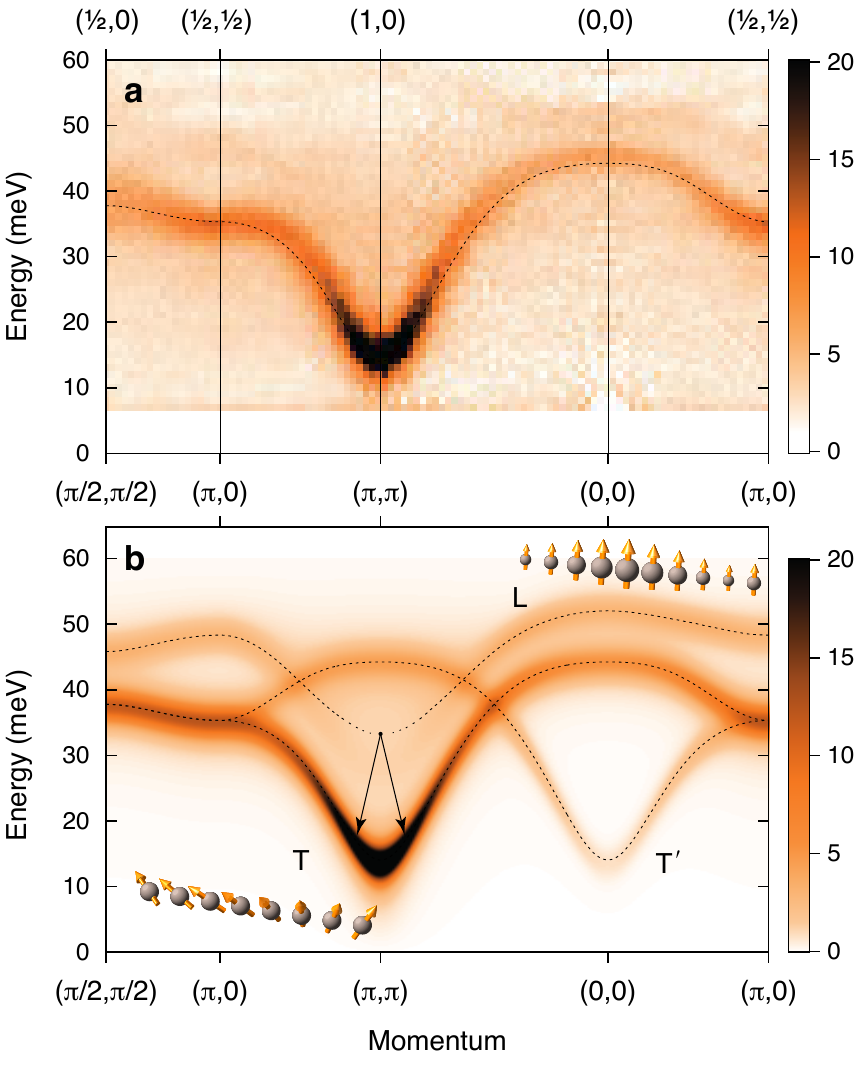}}
\caption{{\bf Spin-wave dispersions strongly deviating from the Heisenberg model.} {\bf a,} TOF INS spectra along high symmetry directions measured at $T$\,=\,5\,K (see Fig.~S3 for more details). The dotted line is from panel {\bf b} for direct comparison between theory and experiment. {\bf b,} The excitation spectra of the model in eq.~(1) calculated with the parameters $E$\,$\simeq$\,25\,meV, $J$\,$\simeq$\,5.8\,meV, $\alpha$\,=\,0.15, $\epsilon$\,$\simeq$\,4.0\,meV, and $A$\,$\simeq$\,2.3\,meV. Transverse and longitudinal modes are labeled as ``T'' and ``L'', respectively, and their motions are depicted. The T$^\prime$ mode arises from back-folding of the T mode by the magnetic ($\pi$,$\pi$) scattering. The L mode carries the Higgs amplitude oscillation. The black arrows show the momentum- and energy-conserving decay process of the L mode into a pair of T modes.}\label{fig:fig1}
\end{figure}

Our comprehensive set of time-of-flight (TOF) inelastic neutron scattering (INS) data over the full Brillouin zone (Fig.~2a) indeed reveal qualitative deviations of the transverse spin-wave dispersion from those of a Heisenberg antiferromagnet. In particular, the global maximum of the dispersion is found at $\mathbf{q}$\,=\,(0,0), in sharp contrast to a Heisenberg antiferromagnet which has a minimum there (Fig.~1). This striking manifestation of orbital magnetism in \CRO\cite{Mizokawa_2001,Haverkort_2008,Fatuzzo_2015} leads us to consider the limit of strong spin-orbit coupling described in terms of a singlet and a triplet separated in energy by $\lambda$ (Fig.~1). In this limit, the ground state is non-magnetic with zero total angular momentum, and therefore a QCP separating it from a magnetically ordered phase is expected as a matter of principle.  Although this QCP can be pre-empted by an insulator-metal transition\cite{Taniguchi_2013,Nakamura_2002} or rendered first-order by coupling to the lattice or other extraneous factors, it is sufficient that the system is reasonably close to the hypothetical QCP.

To assess the proximity to the QCP and the possibility of finding the Higgs mode, we first reproduce the observed transverse spin-wave modes by applying the spin-wave theory\cite{Matsumoto2004,Sommer2001} to the following phenomenological Hamiltonian dictated by general symmetry considerations:
\begin{align}
H &= J\sum_{\langle ij\rangle} (\mathbf{S_{i}\cdot S_{j}}- \alpha S_{zi}S_{zj}) + E \sum_i S^2_{zi} +\epsilon \sum_i S^2_{xi} \notag\\
&\mp A \sum_{\langle ij\rangle} (S_{xi}S_{yj}+S_{yi}S_{xj}).
\end{align}
Here, $\mathbf{S}$ denotes a pseudospin-1 operator describing the entangled spin and orbital degrees of freedom.
This model includes single-ion anisotropy ($E$ and $\epsilon$) terms induced by tetragonal ($z$\,$\parallel$\,$c$) and orthorhombic ($x$\,$\parallel$\,$a$) distortions, correspondingly, as well as an XY-type exchange anisotropy ($\alpha>0$) and the bond-directional pseudodipolar interaction ($A$); note that its sign depends on the bond. Also symmetry allowed---but neglected here---are the Dzyaloshinskii-Moriya interaction (which can be gauged out by a suitable local coordinate transformation) and further-neighbor interactions. The coupling constants resulting from fits of the model to the measured spectra are provided in the caption of Fig.~2. We stress that this model gives the unique minimal description of the system, which we also derive explicitly starting from the microscopic electronic structure (see Supplementary Information). 

\begin{figure}
\centerline{\includegraphics[width=1\columnwidth,angle=0]{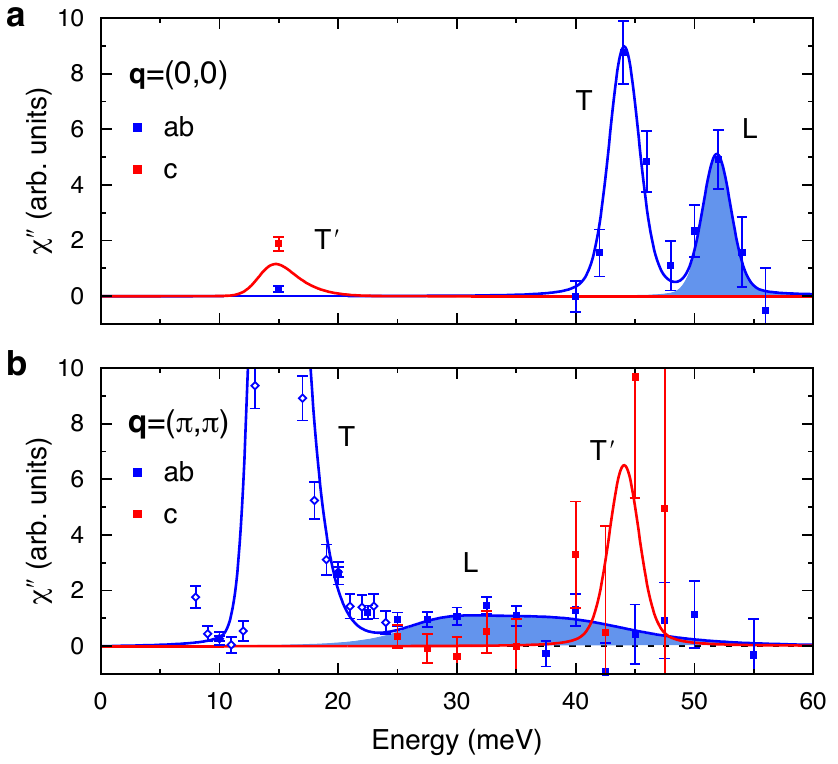}}
\caption{{\bf Identification of the magnetic modes with polarized INS and their comparison to model calculation.} Imaginary part of the dynamic spin susceptibility obtained by normalising the INS spectra measured at $T$\,=\,2\,K at {\bf a,} $\mathbf{q}$\,=\,(0,0) (Fig.~S5) and {\bf b,} $\mathbf{q}$\,=\,($\pi$,$\pi$) (Fig.~S6) with respect to the orientation factor and the isotropic form factor for Ru ion (Fig.~S4). Blue (red) symbols indicate in-plane (out-of-plane) polarized magnetic intensities. Solid symbols show data with the background removed by taking the difference between two spin-flip channels, and open symbols show data from a single spin-flip channel (see Supplementary Information). Error bars denote one standard deviation. 
Solid lines show the calculated spectra, which were convoluted using Gaussian functions with 0.19$\pi$ and 2.5 meV full-width half-maximum to account for the instrumental momentum and energy resolutions, respectively. The decay process of the L mode into T modes is described in the Supplementary Information.  The shaded area indicates the spectral weight of the L mode. The intensities in panels {\bf a} and {\bf b} are in the same arbitrary units.}\label{fig:fig1}
\end{figure}

We find that the single-ion term $E$ 
overwhelms all other coupling constants, particularly the nearest-neighbor exchange coupling $J$, and thus confines the pseudospins to the $ab$ basal plane. This accounts for the XY-like dispersion which has a maximum at $\mathbf{q}$\,=\,(0,0). This important aspect was missed in a recent INS study of \CROns, because the dispersion along the path ($\pi$/2,$\pi$/2)--(0,0) was not measured\cite{Braden_2015}. The large $E$ also acts toward suppressing the magnetic order by favoring the $S_z$\,=\,0 singlet ground state---known in the literature as `spin nematic'\cite{Podolsky2005}---which is also consistent with microscopic considerations (Fig.~1). Other terms play a rather minor role; the pseudodipolar term accounts for the small dispersion along the magnetic zone boundary ($\pi$/2,$\pi$/2)-($\pi$,0), and $\epsilon$ is responsible for gapping the transverse mode, the significance of which will be discussed later.  Our calculation (Fig.~2b) predicts in this parameter regime an intense Higgs mode, visible as a longitudinal spin wave, which heralds a proximate QCP. 

Armed with this specific guidance, we pursue the Higgs mode using spin-polarized INS, using the scattering geometry that maximizes its neutron cross section. We use the standard XYZ-difference method to filter out all non-magnetic and incoherent scattering signals and to resolve all three spin-wave polarizations: the longitudinal mode (L) oscillates along the crystallographic $b$-axis, and the transverse Goldstone modes (T and T$^\prime$) along the $a$ and $c$ axes. Because our sample mosaic consisting of $\sim$100 crystals is ``twinned'', i.e., approximately half of them are rotated 90$^\circ$ about the $c$ axis with respect to the other half, we can only distinguish between in-plane ($ab$) and out-of-plane ($c$) polarized modes. However, this is sufficient to identify the Higgs mode (see Supplementary Information).

Figure 3a shows the measured (symbols with error bars) and calculated (solid lines) dynamical susceptibility at $\mathbf{q}$\,=\,(0,0). We observe three peaks in total as expected, but not all of them were clearly seen in the TOF data because their intensities are maximized in different scattering geometries. The highest-energy peak at $\approx$52 meV is unambiguously identified as the Higgs mode by its magnetic and in-plane-polarized character, because the second in-plane-polarized mode at $\approx$45 meV has already been identified as the T mode (Fig.~2). Further, the data are in excellent accord with the model calculation, which has no adjustable parameter after fitting the dispersion of the T modes. The intensity ratio between the L and T modes is 0.55$\pm$0.11, which is a quantitative measure of the proximity to the QCP (Fig.~S7), at which the distinction between the L and T modes vanishes and their intensities become identical.  

Having established the existence of the Higgs amplitude mode, we now look at its long-wavelength behavior. 
It is at the ordering wave vector where the stability of the Higgs mode critically depends on the dimensionality of the system. In three dimensions, earlier INS studies on a dimerized quantum magnet have established a well-defined Higgs mode\cite{Ruegg_2008}, which was then used to study its critical behavior across a QCP\cite{Ruegg_2008_2,merchant_2014}. In sharp contrast, our in-plane polarized spectrum measured at $\mathbf{q}$\,=\,($\pi$,$\pi$) shows only one clear peak for the T mode at $\approx$14 meV, followed by a broad magnetic intensity distribution in the energy range 20-50 meV, which is however well above the detection limit (Fig.~3b). The Higgs mode has decayed to the extent that a high-flux spin-polarized neutron spectrometer is required to detect its trace.   

\begin{figure}
\vspace{-4mm}
\centerline{\includegraphics[width=1.05\columnwidth,angle=0]{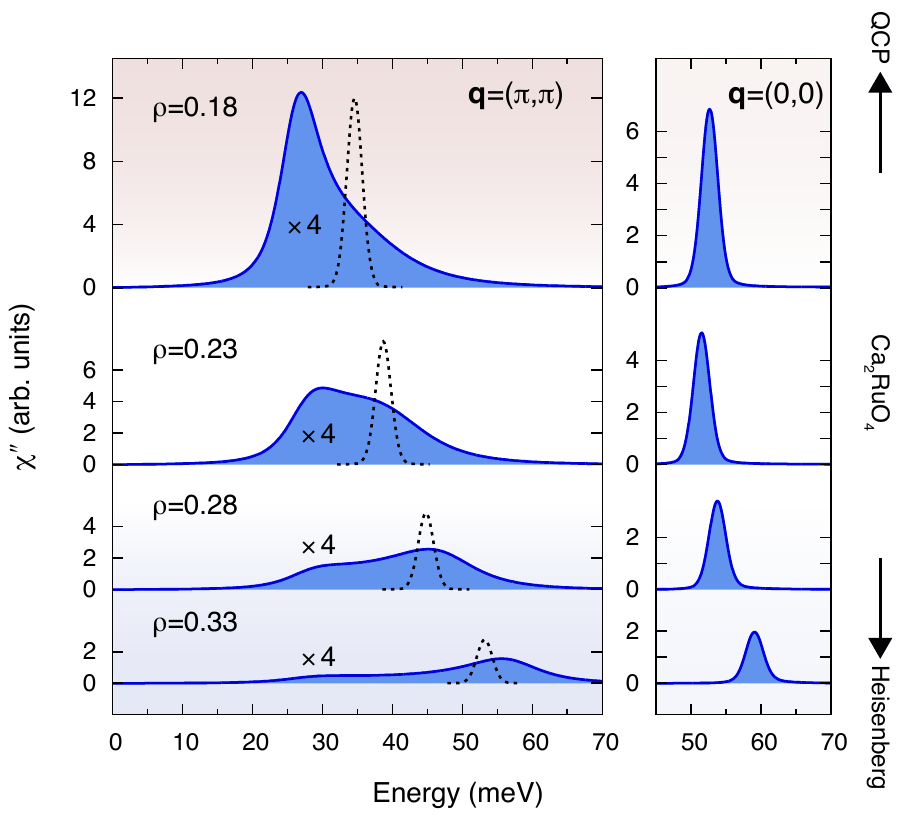}}
\caption{{\bf Evolution of the Higgs mode toward the QCP.} Imaginary part of the longitudinal susceptibility calculated for several values of $J/E$, which can be expressed in terms of the condensate density $\rho$ via the relation $\rho$=$\frac{1}{2}$(1-$E$/8$J$). The $\rho$ values of 0.18, 0.23, 0.28, 0.33 approximately correspond to $J/E$\,=\,0.20,  0.23, 0.28, 0.37, respectively. The dotted lines show the bare susceptibility before taking the decay process into account. }\label{fig:fig1}
\end{figure}

However, it is also known that the response of the Higgs mode strongly depends on the symmetry of the probe being used. Therefore, its rapid decay in the longitudinal susceptibility measured by INS does not necessarily imply its instability in two dimensions. In fact, it has been shown in other two-dimensional systems, such as disordered superconductors\cite{Sherman_2015} and superfluids of cold atoms\cite{Endres_2012}, that the Higgs mode is clearly visible in the scalar susceptibility with its characteristic $\sim$$\omega^3$  onset in the energy spectrum. By contrast, theory predicts that the Higgs mode in the longitudinal susceptibility quickly loses its coherence by decaying into a pair of Goldstone modes\cite{Podolsky_2011,Gazit_2013}. This results in an infrared divergence in two dimensions and renders the Higgs mode elusive.

Conversely, the INS spectrum at $\mathbf{q}$\,=\,($\pi$,$\pi$) encodes detailed information on the decay process of the Higgs mode that is not available from other measurements. To model the decay process, we go beyond the harmonic approximation used in the spin-wave theory to include the coupling of the longitudinal mode to the two-magnon continuum (see Supplementary Information). 
The solid lines in Fig.~3 show the result of the final calculation, which give an excellent description of the data both at $\mathbf{q}$\,=\,(0,0) and $\mathbf{q}$\,=\,($\pi$,$\pi$); the decay process (Fig.~2b) is kinematically restricted away from the ordering wave vector, and the Higgs mode is well identified at $\mathbf{q}$\,=\,(0,0). 

Intriguingly, we encounter a rather unusual situation where all the transverse modes are massive (gapped), as a result of orthorhombic symmetry of the crystal structure parameterized by $\epsilon$. The transverse gap cuts off the infrared singularity and the spectral weight piles up at non-zero energy. We illustrate this point in Fig.~4 by simulating the change in the longitudinal spectrum as the system approaches the QCP. At $\mathbf{q}$\,=\,($\pi$,$\pi$), the decay of the Higgs mode into a pair of minimum-energy transverse modes is still the dominant channel, which generates a `resonance' at twice the energy of the gap. This resonance steals much of the spectral weight from the bare longitudinal mode, thus obscuring its spectral signature especially near the QCP. 
As the system moves away from the QCP, the longitudinal mode progressively hardens and becomes weaker, and its spectral weight spans a larger energy range. The spectral evolution at $\mathbf{q}$\,=\,(0,0) shows this trend with the decay process suppressed; the Higgs mode remains a well-defined excitation even away from the QCP although its intensity quickly diminishes.

Now that we have established a two-dimensional material system, future studies can reveal further aspects of the Higgs mode. In particular, it is uncertain at this point whether the decay process considered above fully describes its dynamics. Other  channels, such as decays into vortex-like excitations, are conceivable in two dimensions and require further investigation. In addition, it would be interesting to compare the results presented herein with the spectra from resonant inelastic x-ray scattering, which can in principle access both the scalar and longitudinal susceptibilities. Finally, it is interesting to note that the Higgs boson in particle physics is detected through its decay products, such as pairs of photons, W and Z bosons, or leptons. The Higgs potential can be determined through the decay rates and branching ratios of these processes, which have been calculated to very high precision. Our study represents the first step toward a parallel development in condensed matter physics.

\vspace {20 pt}
\noindent
{\bf Acknowledgements} We acknowledge financial support from the German Science Foundation (DFG) via the coordinated research program SFB-TRR80, and from the European Research Council via Advanced Grant 669550 (Com4Com). The experiments at Oak Ridge National Laboratory's Spallation Neutron Source were sponsored by the Division of Scientific User Facilities, US DOE Office of Basic Energy Sciences. J.C. was supported by GACR (project no. GJ15-14523Y) and by MSMT CR under NPU II project CEITEC 2020 (LQ1601).

\vspace{20 pt}
\noindent
{\bf Methods}

\vspace{10 pt}
\noindent
{\bf Sample synthesis \& characterization}
Single crystals of \CRO were grown by the floating zone method with RuO$_{2}$ self-flux\cite{Nakatsuji2001}. The lattice parameters $a$\,=\,5.409\,\AA, $b$\,=\,5.505\,\AA, and $c$\,=\,11.9312\,\AA\,  were determined by x-ray powder diffraction, in good agreement with the parameters reported in the literature\cite{Braden_1998} for the ``S'' phase with short $c$-axis lattice parameter. The magnetic ordering temperature $T_\mathrm{N}$\,=\,110\,K was determined using magnetization measurements in a Quantum Design SQUID-VSM device. Polarized neutron diffraction measurements indicate that most of the array orders in the ``A-centered'' magnetic structure with magnetic propagation vector $\mathbf{Q}$\,=\,(1,0,0)\cite{Braden_1998}. The fraction of the sample with ordering vector $\mathbf{Q}$\,=\,(0,1,0), \textit{i.e.}``B-centered'', is estimated to be less than 5\%.

\vspace{10 pt}
\noindent
{\bf Time-of-flight inelastic neutron scattering}
For the TOF measurements, we co-aligned about 100 single crystals with a total mass of $\sim$\,1.5\,g into a mosaic on Al plates. Approximately half of the crystals were rotated 90$^\circ$ about the $c$-axis from the other half (Supplementary Fig.~2). The in-plane and $c$-axis mosaicity of the aligned crystal assembly were  $\lesssim$\,3.2$^\circ$ and  $\lesssim$\,2.7$^\circ$, respectively. The measurements were performed on the ARCS time-of-flight chopper spectrometer at the Spallation Neutron Source, Oak Ridge National Laboratory, Tennessee, USA. The incident neutron energy was 100~meV. The Fermi chopper and  $T_{0}$ chopper frequencies were set to 600 and 90 Hz, respectively, to optimize the neutron flux and energy-resolution. The measurements were carried out at {\it T}\,=\,5\,K. The sample was mounted with ($H$,0,$L$) plane horizontal. The sample was rotated over 90$^\circ$  about the vertical $c$-axis with a step size of 1$^\circ$. At each step data were recorded over a deposited proton charge of 3 Coulombs ($\sim$\,45 minutes) and then converted into 4D $S(\mathbf {Q},\omega)$ using the HORACE software package\cite{Horace} and normalized using a vanadium calibration.

\vspace{10 pt}
\noindent
{\bf Polarized inelastic neutron scattering}
Preliminary triple-axis measurements, in order to reproduce the TOF results and determine the feasibility of the polarized experiment, were done in the thermal triple-axis spectrometer PUMA at the FRM-II, Garching, Germany. The measurements were done on the same sample used for the TOF experiment. To optimize the flux and energy resolution, double-focused PG (002) and Cu (220) monochromators, for measurements below and above 30 meV respectively, and a double-focused PG (002) analyzer were used, keeping $k_{\mathrm f}$\,=\,2.662\,\AA$^{-1}$ constant.
For the polarized triple axis measurement we remounted the crystals from the TOF experiment on Si plates and increased the number of crystals to obtain a total sample mass of $\sim$\,3\,g. The mosaicity of this sample was  $\lesssim$\,3.2$^\circ$ and  $\lesssim$\,2.6$^\circ$ for in-plane and $c$-axis, respectively. The experiment was performed on the IN20 three-axis-spectrometer at the Institute Laue-Langevin, Grenoble, France. For the XYZ polarization analysis, we used a Heusler (111) monochromator and analyzer in combination with Helmholtz coils at the sample position. Throughout the experiment we used a fixed $k_{\mathrm f}$\,=\,2.662\,\AA$^{-1}$ and performed polarization analysis in energy and $H$ scans at ($\pi$,$\pi$) and (0,0), keeping $L$ as small as permitted by kinematic constraints. The measurements were carried out at {\it T}\,=\,2\,K.
 
\bibliography{biblio.bib}


\newpage
\noindent 
{\bf Supplementary Information}

\vspace{20 pt}
\noindent
{\bf A. Microscopic model}

\vspace{10 pt}
\noindent

\begin{figure}[b]
\includegraphics[width=0.9\columnwidth]{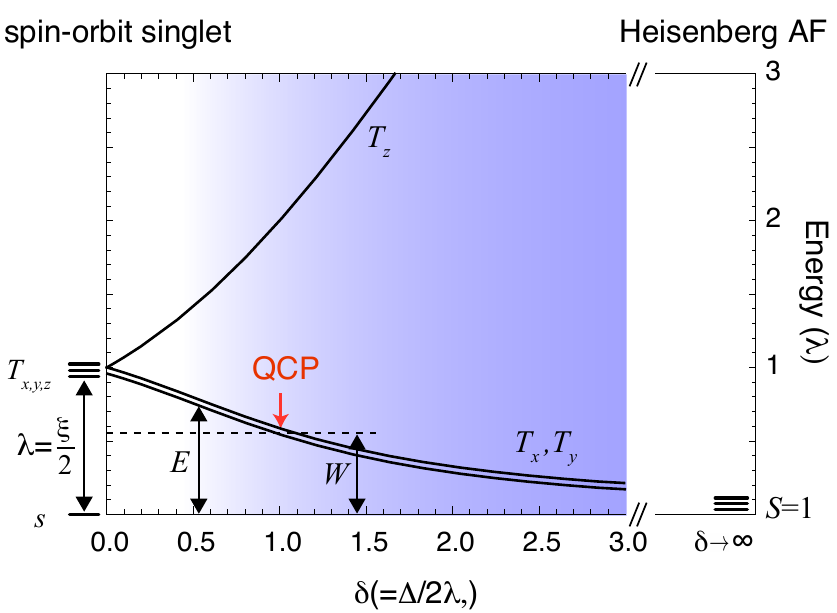}
\setcounter{figure}{0}
\renewcommand{\thefigure}{S\arabic{figure}}
\caption{{\bf Microscopic mechanism of the QCP driven by tetragonal lattice distortion.} Crystal-field splitting of the triplet effectively lowers the energy scale of SOC from $\lambda$ to $E$. $\lambda$ is equal to one-half of the single-electron SOC $\xi$ for $d^4$ low-spin electron configuration. The QCP occurs when $E$ becomes equal to the strength of the exchange field $W$\,$\approx$\,2$zJ$. The blue shading indicates the region where the effective $S$\,=\,1 model is valid. }
\end{figure}

\begin{figure}[b]
\includegraphics[width=1.0\columnwidth]{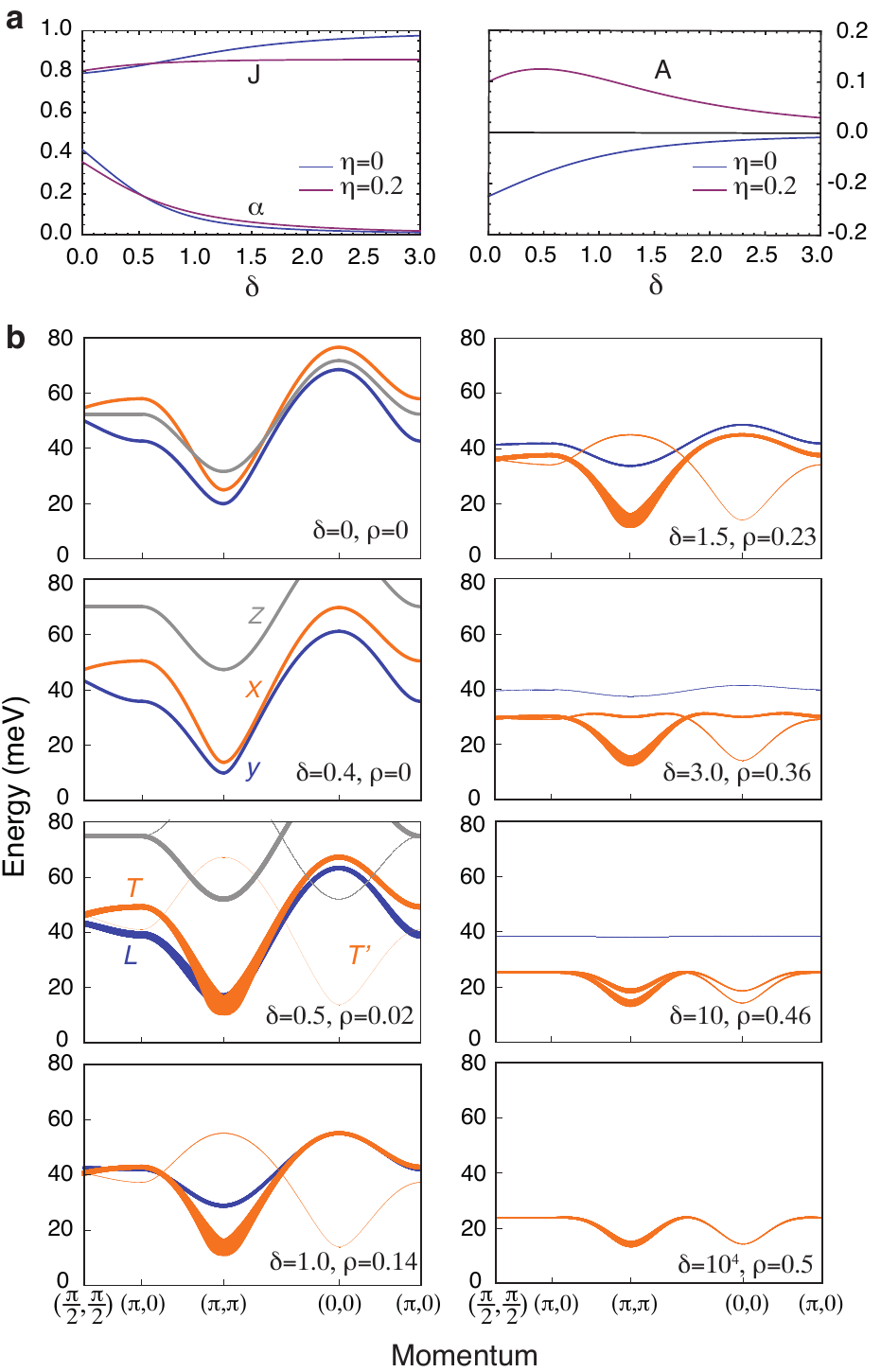}
\renewcommand{\thefigure}{S\arabic{figure}}
\caption{{\bf Coupling constants and spin-wave dispersion from the microscopic model.} {\bf a,} $J$ and $A$ (see eq.~1) are in units of $t^2/U$. {\bf b,} Evolution of the spin-wave dispersions from the non-magnetic spin-orbit singlet to the Heisenberg limit as a function of $\delta$. $\eta$\,=\,0.25 was used. A non-zero $\rho$, the condensate density, indicates the magnetic order. At $\delta$\,=\,0.5, the magnetic order has set in; now the $T_x$ becomes the transverse mode and $T_y$ the longitudinal mode. At $\delta$\,=\,1.5, the simulation is very similar to the experimental data. Further increase of $\delta$ leads to a development of a local minimum at $\mathbf{q}$\,=\,(0,0), and eventually to a Heisenberg-like dispersion.}\label{fig:fig1}
\end{figure}

We derive here the phenomenological model in eq.~(1) starting from the microscopic electronic structure. The compressive tetragonal distortion $\Delta$, is the key parameter that determines the proximity of \CRO to the QCP, because the spin-orbit splitting $\lambda$ is known for Ru(IV) ion, and the nearest-neighbor exchange coupling $J$ is to a large extent fixed by the measured bandwidth $W$\,$\simeq$\,2$zJ$ of the spin-wave ($z$ is the coordination number). In the absence of $\Delta$, the low energy physics is described in terms of a singlet-triplet model\cite{Khaliullin_2013}, formally similar to that used for dimerized quantum magnets\cite{Ruegg_2008_2}, such as  TlCuCl$_3$ and BaCuSi$_2$O$_6$. The magnetic transitions in these systems have been extensively studied as a Bose-Einstein condensation of triplons, where the magnetic field $H$ plays the role of the chemical potential $\mu$. In our case, $\Delta$ plays the role of $\mu$ (see Fig.~S1); it splits the triplets into a doublet and a singlet and thereby lowers the energy cost $E$ to create an exciton ($T_x$ or $T_y$). The quantum phase transition occurs when $E$\,$\simeq$\,$W$; the equality holds for classical consideration. With the free-ion value $\lambda$\,$\approx$\,75 meV and $W$\,$\approx$\,45 meV, we estimate that QCP is at $\delta$($\equiv$$\Delta$/$2 \lambda$)\,$\approx$\,1. 

\begin{figure*}
\includegraphics[width=1.55\columnwidth]{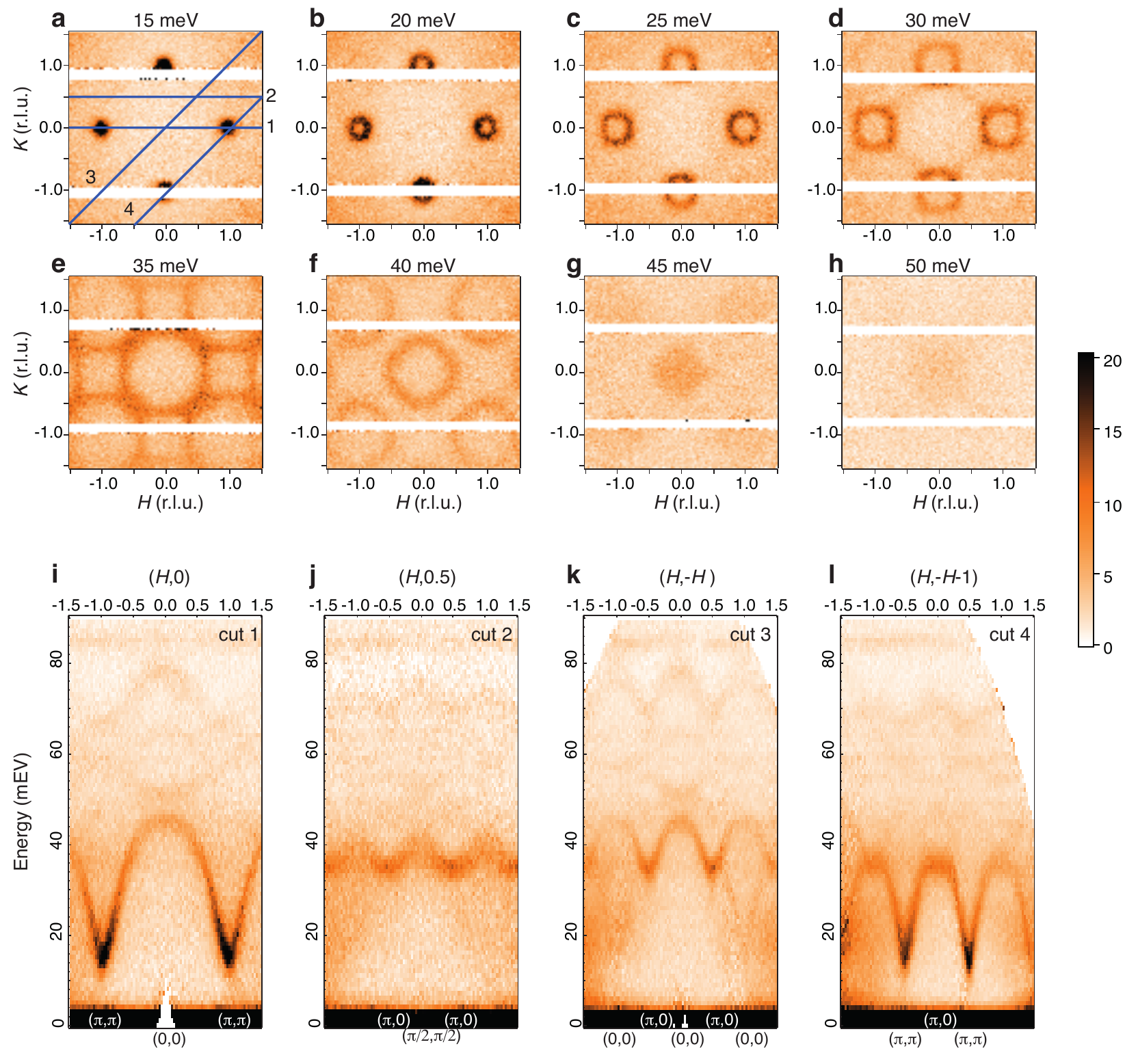}
\renewcommand{\thefigure}{S\arabic{figure}}
\caption{{\bf Time-of-flight INS spectra of \CROns.}  {\bf a-h,} Constant energy maps of INS intensity in the ($H$,$K$) plane with $L$ integrated. The wave-vectors are expressed in reciprocal lattice units (r.l.u.) of a tetragonal cell with  $a^{*}$=\,$b^{*}$=\,1.16\,\AA$^{-1}$ and $c^{*}$=\,0.53\,\AA$^{-1}$. At $E$\,=\,14\,meV, intense spin waves are observed at the reduced wave-vectors $(\pm 1,0)$ and $(0,\pm 1)$, which correspond to the antiferromagnetic ordering wave-vector $(\pi,\pi)$ for the 2D square lattice Brillouin-zone.  {\bf i-l,} Energy spectra along high-symmetry directions as shown in blue lines in panel {\bf a}. The intensity is in arbitrary units. }
\end{figure*}

Because \CRO is on the right hand side of the QCP where the $T_z$ singlet is in very high energy and hence can be neglected, the low-energy physics of \CRO can be described by the three levels $\{s,T_x,T_y\}$. These three levels constitute the effective $S$\,=\,1 degrees of freedom in the phenomenological model in eq.~(1). We note that the large energy scale of $E$ microscopically originates from $\lambda$ and depends on $\delta$.  

In terms of the $z$ projections of the $S$=1 and $L$=1 
moments $\lvert S_M, L_M\rangle$, the wave functions for the basis states $\{s,T_x,T_y\}$ with 
$T_x=\frac1{i\sqrt2}(T_1-T_{-1})$, $T_y=\frac1{\sqrt2}(T_1+T_{-1})$ are given by 
\begin{align}
&|s\rangle = \sin\theta_0 \,\tfrac1{\sqrt2}(|1,-1\rangle+|-1,1\rangle)
           -\cos\theta_0 \,|0,0\rangle \;, \\
&|T_{+1}\rangle = \cos\theta_1 |1,0\rangle - \sin\theta_1 |0,1\rangle \;, \\
&|T_{-1}\rangle = \sin\theta_1 |0,-1\rangle - \cos\theta_1 |-1,0\rangle.
\end{align}
Here the angles $\theta_0$, $\theta_1$ are defined through
\begin{equation}
\tan\theta_1 = \frac1{\delta+\sqrt{1+\delta^2}} \;, \quad
\tan\theta_0 = \sqrt{1+\beta^2}-\beta, 
\end{equation}
where $\beta=\frac1{\sqrt2}(\delta-\frac12)$. 
The energy $E$ in eq.~(1) and Fig.~S1 is given by
\begin{equation}
E=\frac{\xi}{2} \left(\frac{\sqrt{2}}{\beta+\sqrt{1+\beta^2}}-\frac{1}{\delta+\sqrt{1+\delta^2}}\right).
\end{equation}

Using the above wavefunctions in the standard second-order perturbation theory, we calculate the coupling constants incorporating the Hund's coupling $\eta$\,=\,$J_H$/$U$ measured in units of the Coulomb interaction $U$. Figure S2a shows the exchange constant $J$, two-ion XY-type anisotropy $\alpha$, and pseudodipolar interaction $A$ as functions of $\delta$,
for $\eta$\,=\,0 and $\eta$\,=\,0.2; the latter would be more realistic. The calculation shows
that $\alpha$\,$\ll$\,1 in the entire range of $\delta$ where the model is relevant to \CROns, insensitive to the value
of $\eta$,
confirming that the XY-type anisotropy due to two-ion exchange is small. Thus, the single-ion term $E$ is mostly responsible for the XY-type anisotropy.

Using these coupling constants, we simulate in Fig.~S2b the evolution of the spin-wave spectra over the entire phase diagram from the non-magnetic singlet to the Heisenberg limit using $\delta$ as the only tuning parameter. In the simulation, we used $\lambda$\,=\,50 meV, $\eta$\,=\,0.25, and $t^2/U$\,=\,5.75 meV. Additionally the $\epsilon$ term in eq.~(1) was added to reproduce the transverse mode gap of $\approx$14 meV. Note that at $\delta$\,=\,1.5, the simulated spectra becomes very similar to the experimental spectra. The above parameters translate to $J$\,$\simeq$\,5.2\,meV, $\alpha$\,=\,0.1, $E$\,=\,21.5\,meV and $A$\,$\simeq$\,1.0\,meV, which are in excellent agreement with those found from the fitting, considering that the model is minimal and the coupling constants absorb various renormalization effects in the solid not taken into account in the microscopic model. In particular, $A$ absorbs the effect of further neighboring couplings, which also contribute to the dispersion along the magnetic zone boundary ($\pi$/2,$\pi$/2)-($\pi$,0).

\vspace{10 pt}
\noindent
{\bf B. Time-of-flight inelastic neutron scattering}

Figures S3a-h and S3i-l exhibit constant-energy maps and energy spectra along high-symmetry directions, respectively, measured by TOF INS comprising spin-wave dispersions in the energy range 14\,$\lesssim$\,$\hbar\omega$\,$\lesssim$\,45 meV and phonon modes above $\sim$50\,meV. The magnetic nature of the former is explicitly confirmed by using spin-polarized neutrons (Fig.~3), and the non-magnetic nature of the latter is inferred from exhaustion of all magnetic modes and also through comparison with the known phonon modes. The data has been integrated along $L$ because the magnetic excitations are close to the 2D limit, which can also be seen from the narrow linewidth of the excitations after the integration. Indeed, a recent INS study shows that the excitations at $(\pi,\pi)$ are almost dispersionless (less than 1\,meV), with no significant change in amplitude\cite{Braden_2015}. 

\vspace{10 pt}
\noindent
{\bf C. Polarization analysis}

\begin{figure}
\renewcommand{\thefigure}{S\arabic{figure}}
  \centering
  \includegraphics[width=0.85\columnwidth]{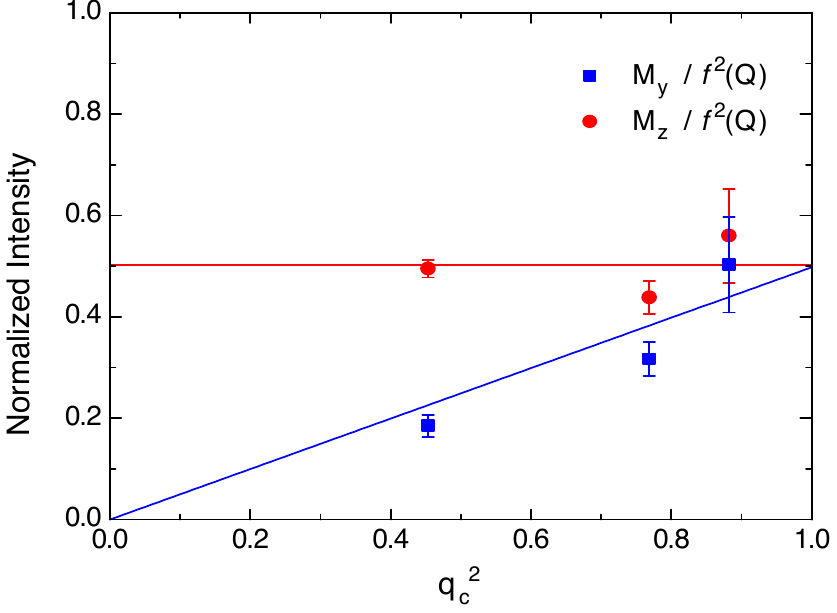}

  \caption{{\bf Determination of the twinning ratio.} Magnetic intensities $M_y$(blue squares) and $M_z$(red circles) normalized by the squared magnetic form factor as a function of $q_c^2$ at $\mathbf{Q}$\,=\,(1,0,$L$) with $L$\,=\,2, 4, and 6 at energy transfer 15\,meV. For one type of domain the intensity is constant, and for the other type of domain the intensity increases linearly with $q_c^2$. A one-parameter fit (red and blue solid lines) to the data points determines the twinning ratio $p$.}
\label{Fig:twin_ratio}
\end{figure}

\vspace{10 pt}
\noindent
In the standard reference frame for the neutron polarization with $\hat{x}\parallel\mathbf{Q}$, $\hat{y}\perp\mathbf{Q}$ in the scattering plane of the spectrometer and $\hat{z}=\hat{x}\times\hat{y}$, the magnetic intensity in the spin flip channels is extracted from the differences:
%
%
\begin{align}
  M_{y} &= I_x - I_y, \\ \nonumber
  M_{z} &= I_x - I_z,
\end{align}
where $I_x$, $I_y$, $I_z$ are the raw intensities of the respective polarizations. Note that any contribution from the background is suppressed in the difference. For conversion from INS intensity to dynamic spin susceptibility, we used the isotropic form factor for Ru$^+$, which gave a good description of the data at 15 meV (Fig.~\ref{Fig:twin_ratio}). 

\vspace{10 pt}
\noindent
{\bf C.1 Twinning ratio}

\vspace{10 pt}
\noindent
In this study the $a$ and $b$ orientation of the crystals in the array are not distinguished. In other words, for the volume fraction $p$ of the sample the scattering plane is ($H$,0,$L$), and for the fraction $(1-p)$ the scattering plane is (0,$H$,$L$). Taking into account the polarization factor, the intensities in each channel are related to excitations $M_{a}$, $M_{b}$ and $M_{c}$ along the crystallographic directions by:
\begin{align}
 M_{y} &= q_c^2 \left[ p~M_{a} + \left(1-p \right) M_{b} \right] + \left(1-q_c^2\right) M_{c}\\ \nonumber
 M_{z} &= \left(1-p \right) M_{a} + p~M_{b}
\end{align}
where $q_c^2$\,=\,$\left(Q_c/\lvert\mathbf{Q}\rvert\right)^2$.

The twinning ratio $p$ can be estimated from rocking scans through the Bragg reflections (4,0,0) and (0,4,0) where the separation in the scattering angle is large enough to distinguish the two peaks (not shown). Alternatively, $p$ can be estimated from the inelastic measurements by considering the $L$-dependence of the 15\,meV feature at ($\pi$,$\pi$) as shown in Fig.~\ref{Fig:twin_ratio}. Since this is an in-plane transverse mode, $M_b$ and $M_c$ vanish and eq.~(14) greatly simplifies. From the one-parameter fit to the data, a twinning ratio $p$\,=\,0.498$\,\pm\,0.014$ is determined, consistent with the first method. For the analysis we used $p$\,=\,0.5. 

From polarization analysis on this ``twinned'' array, only in-plane (ab) or out-of-plane (c) polarization can be distinguished, as $M_{a}$ and $M_{b}$ give equal contributions in each channel. Nevertheless, two in-plane-polarized modes T and L and one out-of-plane-polarized mode T$^\prime$  are expected which are non-degenerate. We can indeed distinguish them from the energy scans at (0,0) and $(\pi,\pi)$.

\vspace{10 pt}
\noindent
{\bf C.2 Energy scans at $\mathbf{q}$\,=\,(0,0)}
\vspace{10 pt}

\noindent
For the energy scans at $\mathbf{q}$\,=\,(0,0), it is useful to use two different Brillouin zones to maximize the intensity for the different modes. The measurements at $\mathbf{Q}$\,=\,(2,0,0.4) at 15\,meV give conclusive evidence for the folded mode T$^\prime$ (Fig.~S5a), as the out-of-plane polarization gives rise to a signal in $M_y$ but not in $M_z$. We confirmed that the signal is peaked at (2,0,0.4) by scanning along the $H$ direction (not shown). To avoid a sharp spurion a small $L$ component was used. 
For the energy scan at $\mathbf{Q}$\,=\,(0,0,$L$), shown in Fig.~S5b, the signal exclusively originates from in-plane polarized modes. We observe two magnetic excitations clearly separated in energy, both with equal contributions from the $M_{y}$ and $M_{z}$ channels. Given the dispersion obtained from TOF, the peak at 45 meV is unambiguously assigned to the transverse mode T; thus the peak at 52\,meV must be associated with the longitudinal mode L.
\begin{figure}
  \centering
  \includegraphics[width=0.9\columnwidth]{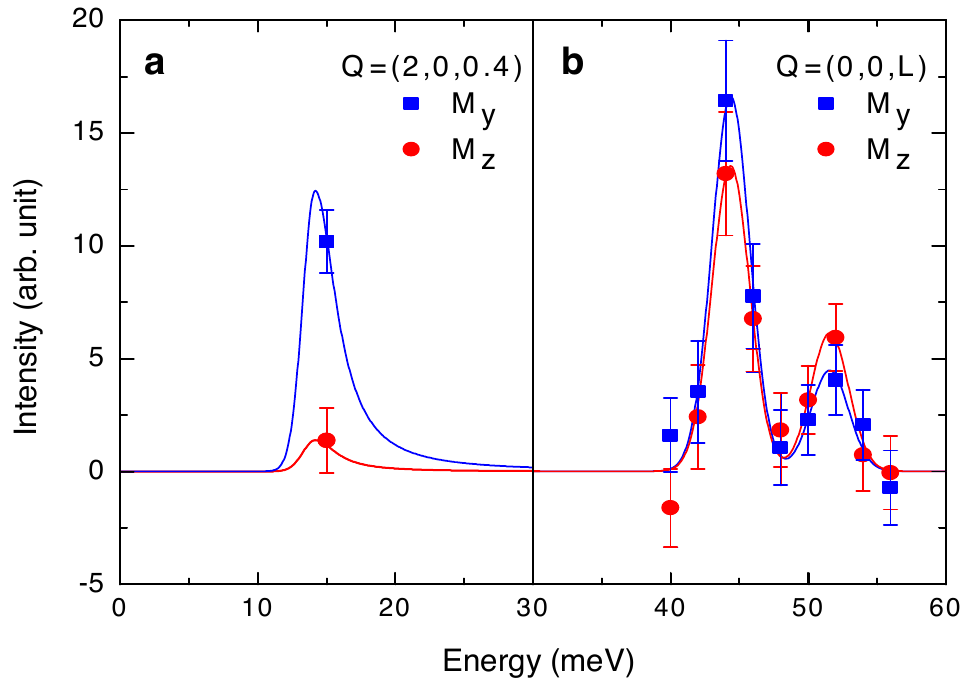}
\renewcommand{\thefigure}{S\arabic{figure}}
  \caption{{\bf Energy scans at $\mathbf{q}$\,=\,(0,0).} Magnetic intensities {\bf a,} at $\mathbf{Q}=(2,0,0.4)$, and {\bf b,} at $\mathbf{Q}=(0,0,L)$. The value of $L$ was varied along the scan to minimize the magnitude of $\mathbf{Q}$. Blue squares denote $M_y$, red circles $M_z$, and the lines are guides to the eye.}

\end{figure}

\vspace{10 pt}
\noindent
{\bf C.3 Energy scans at $\mathbf{q}$\,=\,$(\pi,\pi)$}
\vspace{10 pt}

\noindent
Energy scans at $\mathbf{Q}$\,=\,($1$,0,$L$) shown in Fig.~S6, corresponding to  $\mathbf{q}$\,=\,$(\pi,\pi)$ of the tetragonal unit cell, reveal three magnetic excitations above a gap of 14\,meV. For the two features lowest in energy we observe a signal in both $M_y$ and $M_z$ channels, characterizing them as in-plane polarized magnetic excitations. 
The third mode is unambiguously identified as the folded mode as the polarization factor suppresses the intensity in the $M_{z}$ channel completely for out-of-plane excitations. The $M_y$ and $M_z$ signals allow separation of the in-plane and out-of-plane responses, because the $M_y$ signal exclusively originates from in-plane polarized modes, whereas the $M_z$ signal has contributions from both in-plane and out-of-plane polarized modes.

\begin{figure}
  \centering
  \includegraphics[width=0.9\columnwidth]{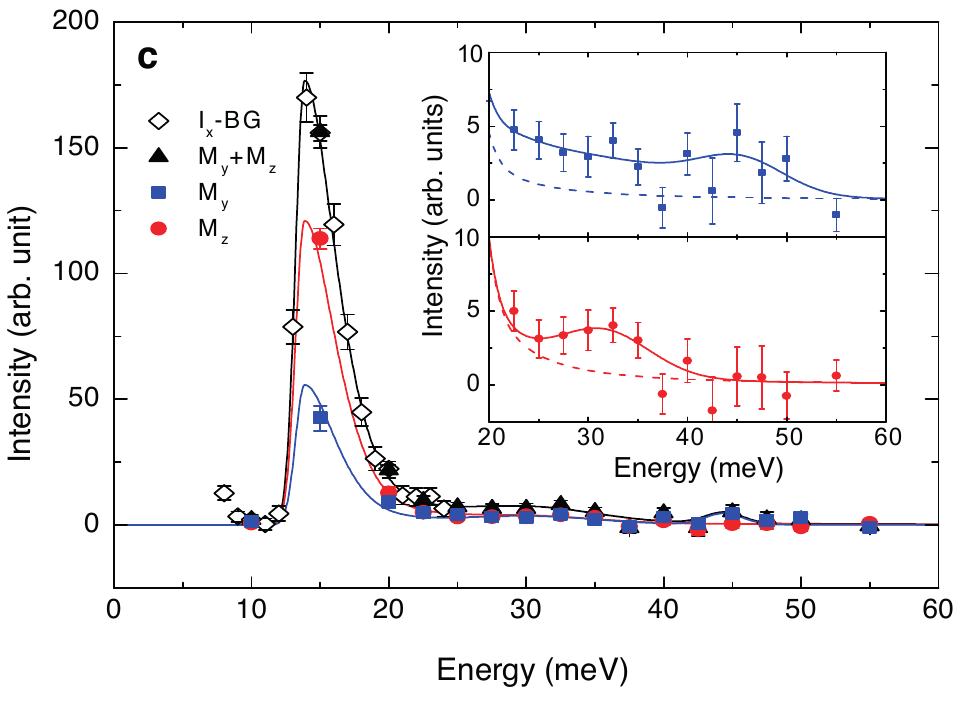}
\renewcommand{\thefigure}{S\arabic{figure}}
  \caption{{\bf Energy scan at $\mathbf{q}$\,=\,($\pi$,$\pi$).} The value of $L$ was varied along the scan to minimize the magnitude of $\mathbf{Q}$. The data denoted $I_x$-BG (empty black diamonds) is obtained from the raw data in the $M_x$ channel after subtraction of a small background; this method is only reliable when the signal is much larger than the background. The intensities $M_y$+$M_z$(filled black triangles), $M_y$ (blue squares), and $M_z$(red circles) are obtained using eq.~(13) and the lines are guides to the eye. The inset shows in detail the region above 20\,meV for $M_y$ (top) and $M_z$(bottom). Dashed lines represent the tail of the main transverse mode.}
\label{Fig:pipi}
\end{figure}

\vspace{10 pt}
\noindent
{\bf D. Proximity to the QCP}

To quantify the proximity to the QCP, we introduce $\tau$\,=\,$J/J_{\textrm cr}$\,$\approx$\,$8J/E$. At the QCP ($\tau$\,=\,1), where the distinction between transverse and amplitude modes vanishes, their intensity ratio at $\mathbf{q}$\,=\,(0,0) is $\simeq$\,1 (equality holds when the gap is zero), and approaches zero as the moment saturates (Fig.~S7). The measured intensity ratio of 0.55\,$\pm$\,0.11 translates to $\tau$\,$\approx$\,1.8. In principle, the size of the static moment contains the same information, but only after corrections due to $g$-factors, covalency, and quantum fluctuations, have been properly taken into account, which are model-dependent and fraught with systematic uncertainties. 

\begin{figure}[b]
  \centering
  \includegraphics[width=0.9\columnwidth]{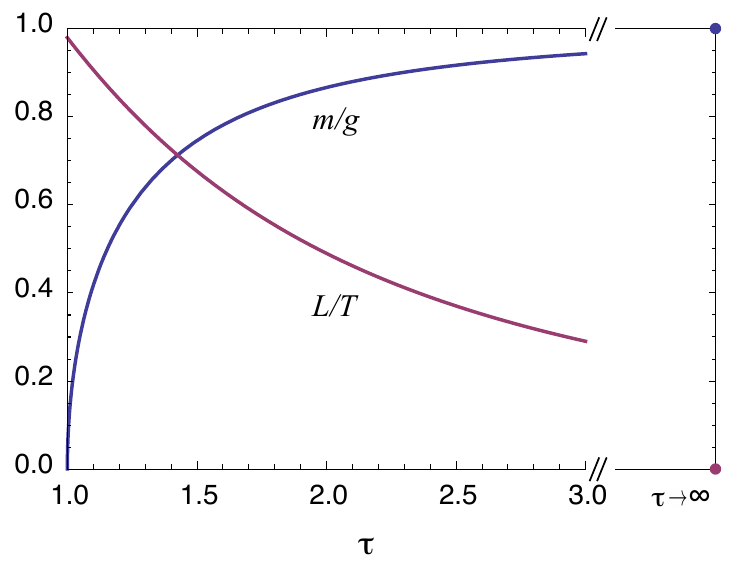}
\renewcommand{\thefigure}{S\arabic{figure}}
  \caption{{\bf Quantification of the proximity to the QCP.} Evolution of the intensity ratio between the amplitude (L) and the transverse (T) modes at $\mathbf{q}$\,=\,(0,0), and the static magnetic moment normalized by the in-plane $g$-factor as a function of $\tau$\,=\,$J/J_\mathrm{cr}$. $\tau$\,=\,1 at the QCP.}
\label{Fig:pipi}
\end{figure}

\vspace{10 pt}
\noindent
{\bf E. Mode dispersions and intensities}

\vspace{10 pt}



\noindent
The excitation spectra for the model in eq.~(1) formulated in the basis
$\{s,T_x,T_y\}$ were calculated using the modified spin-wave theory for
the models where the QCP is associated with triplet 
condensation (see refs.~19,20).
The energy and magnetic intensity of the longitudinal mode 
obtained within the harmonic approximation reads as
\begin{equation}
\omega_{L\vc q} = W\sqrt{1+\frac{\gamma_{\vc q}}{\tau^2}} \;,\quad
I_{L\vc q} \propto \frac{1}{\tau}\frac1{\sqrt{\tau^2+\gamma_{\vc q}}} \;,
\end{equation}
where $W$\,=\,$8J$ is the energy scale and 
$\gamma_{\vc q}=\frac12(\cos q_x+\cos q_y)$. 
This mode is most dispersive and intense
near the QCP ($\tau$\,$\sim$\,1) while in the rigid-spin limit
($\tau$\,$\gg$\,1), it 
flattens and vanishes. To describe the main (T) and folded (T$^\prime$)
transverse modes, we introduce two auxiliary quantities
\begin{align}
a_{\vc q}&=\tfrac12 W(1+\tfrac1{\tau})(1+\gamma_{\vc q})+\epsilon\;,\notag\\
b_{\vc q}&=\tfrac12 W(1+\tfrac1{\tau})\left[
1-\tfrac{\tau-1}{\tau+1}(1-\alpha)\gamma_{\vc q} \right]+\epsilon \;.
\end{align}
Then the energy and intensity of the T mode may be expressed as
\begin{equation}
\omega_{T\vc q} = \sqrt{a_{\vc q} b_{\vc q}} \;,\quad
I_{T\vc q} \propto \frac{1}{\tau} \frac{\tau+1}2 
\sqrt{\frac{b_{\vc q}}{a_{\vc q}}}
\end{equation}
and for the T$^\prime$ mode we have
\begin{equation}
\omega_{T'\vc q} = \omega_{T\vc{\tilde q}} \;,\quad
I_{T'\vc q} \propto \frac{1}{\tau} \frac{\tau-1}2 
\sqrt{\frac{a_{\vc{\tilde q}}}{b_{\vc{\tilde q}}}} \;,
\end{equation}
where $\vc{\tilde q} = \vc q + (\pi,\pi)$. 
The intensity contrast between the 
T and T$^\prime$ modes is most pronounced in the soft-spin situation
where $\tau$\,$\sim$\,1. At the crossing point of their dispersions, 
$\vc q$\,=\,$(\frac\pi2,\frac\pi2)$, $\gamma_{\vc q}$ is zero and $I_{T'}/I_T$
becomes $(\tau-1)/(\tau+1)$, vanishing as $\tau$\,$\rightarrow$\,1. Note that in the standard Heisenberg or XY models the intensity ratio is 1 (consider $\tau$\,$\rightarrow$\,$\infty$). For the relative intensity 
of the L and T mode at $\vc q$\,=\,$(0,0)$, used to quantify the proximity 
to the QCP, we get $I_L/I_T=2/\sqrt{(\tau+1)(\tau^2+1)}$ ($\alpha$\,=\,$0$, 
$\epsilon$\,=\,0) corrected by a multiplicative factor 
$1-\frac{\tau^2}{2(\tau+1)}\frac{\epsilon}{W}$ for small nonzero $\epsilon$.
Non-zero pseudodipolar term in eq.~(1) mixes the L mode and T mode 
leading to two modes with the modified dispersions
\begin{equation}
\omega_{1,2\vc q}^2 = \frac{\omega_{L\vc q}^2+\omega_{T\vc q}^2}2 
\pm \sqrt{ \left(\frac{\omega_{L\vc q}^2-\omega_{T\vc q}^2}2\right)^2 
+ c_{\vc q}^2 } \;,
\end{equation}
where $c_{\vc q}^2=W^3 b_{\vc q} (A/2J\tau)^2 (\cos q_x-\cos q_y)^2$.
Due to the $d$-wave type form-factor $(\cos q_x-\cos q_y)$, this 
correction is only relevant near the $(\pi,0)$ area.

The two-dimensional situation requires us to go beyond the harmonic
approximation for the amplitude mode. Its coupling to the two-magnon continuum
modifies the bare susceptibility
\begin{equation}
\chi_{L0}(\vc q,\omega)=\frac{W}{2(\omega_{L\vc q}^2-\omega^2)}
\end{equation}
associated with the amplitude mode as $\chi_L^{-1}$\,=\,$\chi_{L0}^{-1}-\Pi_L$. 
Collecting the leading terms, the self-energy $\Pi$ is obtained as
\begin{equation}
\Pi_L(\vc q,\omega) = \sum_{\vc k}
\frac{ M^2_{L\vc k\vc k'} b_{\vc k}b_{\vc k'} 
(\omega^{-1}_{T\vc k}+\omega^{-1}_{T\vc k'}) }
{(\omega_{T\vc k}+\omega_{T\vc k'})^2-(\omega+i\Gamma)^2} \;.
\end{equation}
Here $\vc k'=-\vc k+\vc q+(\pi,\pi)$ and the matrix element
\begin{equation}
M^2_{L\vc k\vc k'} = \frac{W^2}4 \left(1-\frac1{\tau^2}\right) 
\left( \frac{\gamma_{\vc q}}{\tau} 
      +\frac{\gamma_{\vc k}+\gamma_{\vc k'}}2 \right)^2 \;.
\end{equation}
In the calculations, we have used the broadening parameter 
$\Gamma$\,=\,$6\:\mathrm{meV}$.
The self-energy is largest for $\vc q$\,$\approx$\,$(\pi,\pi)$, where the dominant
contribution comes from $\vc k\approx -\vc k' \approx (\pi,\pi)$ (supported
by both small $\omega_T$ and large $M^2_{\vc k\vc k'}$), and turns 
the amplitude mode into a broad feature. 
A sizable gap $\omega_{T(\pi,\pi)}$ of the spin-wave dispersion prevents the infrared
singularity of $\Pi$, whose imaginary part would diverge like $1/\omega$ in
the gapless case. In our case it is zero below the cutoff energy 
$2\omega_{T(\pi,\pi)}$ comparable to $W$ making the above perturbative
approach well controlled.

The T mode is the subject of a similar, albeit much smaller
renormalization due to a coupling to the L mode. 
The relevant bare susceptibility
$\chi_{T0}(\vc q,\omega)=\frac12 b_{\vc q}/(\omega_{T\vc q}^2-\omega^2)$
is modified according to $\chi_T^{-1}$\,=\,$\chi_{T0}^{-1}-\Pi_T$
by employing the self-energy
\begin{equation}
\Pi_T(\vc q,\omega) = \sum_{\vc k}
\frac{ M^2_{T\vc k\vc k'} W b_{\vc k'} 
(\omega^{-1}_{L\vc k}+\omega^{-1}_{T\vc k'}) }
{(\omega_{L\vc k}+\omega_{T\vc k'})^2-(\omega+i\Gamma)^2} 
\end{equation}
containing the matrix element
\begin{equation}
M^2_{T\vc k\vc k'}= \frac{W^2}8 \frac{\tau+1}{\tau^2(\tau-1)}
\left(1\!+\!\gamma_{\vc q}\!-\!\gamma_{\vc k}\!-\!\gamma_{\vc q-\vc k} 
\!+\! 2\tau \frac{\epsilon}W \right)^2 \;.
\end{equation}	
Note that, as a consequence of the rotational symmetry, this coupling vanishes
for the $\vc q$\,=\,($\pi$,$\pi$) magnons in the gapless situation ($\epsilon$\,=\,0).





\end{document}